\documentclass[manuscript=article]{achemso}

\usepackage{amsmath,amssymb}
\usepackage{graphicx}
\usepackage{booktabs}
\usepackage{braket}       % Dirac notation
\usepackage{xcolor}
\usepackage{algorithm}
\usepackage{algpseudocode}
\algnewcommand\algorithmicforeach{\textbf{for each}}
\algdef{S}[FOR]{ForEach}[1]{\algorithmicforeach\ #1\ \algorithmicdo}
\usepackage{hyperref}


\title{Automated Unitary Coupled Cluster Circuit Design via
       Differentiable Quantum Architecture Search}

\author{Jianpeng Chen}
\affiliation{Guangdong Basic Research Center of Excellence for Aggregate
             Science, School of Science and Engineering, The Chinese
             University of Hong Kong (Shenzhen), Shenzhen, Guangdong
             518172, P. R. China}

\author{Zirui Sheng}
\affiliation{Guangdong Basic Research Center of Excellence for Aggregate
             Science, School of Science and Engineering, The Chinese
             University of Hong Kong (Shenzhen), Shenzhen, Guangdong
             518172, P. R. China}

\author{Cunxi Gong}
\affiliation{Guangdong Basic Research Center of Excellence for Aggregate
             Science, School of Science and Engineering, The Chinese
             University of Hong Kong (Shenzhen), Shenzhen, Guangdong
             518172, P. R. China}

\author{Weitang Li}
\email{liweitang@cuhk.edu.cn}
\affiliation{Guangdong Basic Research Center of Excellence for Aggregate
             Science, School of Science and Engineering, The Chinese
             University of Hong Kong (Shenzhen), Shenzhen, Guangdong
             518172, P. R. China}


\begin{document}
% ===================================================================

%% ----- Abstract ---------------------------------------------------
\begin{abstract}
Designing compact and accurate circuits for the variational quantum
eigensolver (VQE) is a central challenge in near-term quantum chemistry.
Existing adaptive methods such as ADAPT-VQE design circuits by iteratively
selecting operators from a predefined pool guided by gradient information
and greedy heuristics.
In this work, we adopt differentiable quantum architecture search (DQAS)
as a circuit design framework based on the UCCSD operator pool, and
introduce two complementary strategies: a global mode that simultaneously
optimizes all operator selections, and a layerwise mode that constructs
circuits incrementally while preserving previously learned structure.
By relaxing discrete operator selection into a continuous differentiable
optimization, DQAS enables gradient-based exploration over the
combinatorial space of UCC circuit architectures.
Benchmarks on BeH$_2$, H$_4$, LiH, H$_6$, and H$_2$O (8--14 qubits)
show that both strategies achieve higher accuracy and fewer CNOT gates
than ADAPT-VQE in the compact circuit regime, with up to 2.7-fold
accuracy improvement for H$_2$O and CNOT reductions of 13--17\% at
equivalent circuit depths.
Benchmarks on the qubit-excitation-based (QEB) operator pool confirm
that both advantages generalize beyond UCCSD.
These results demonstrate that differentiable architecture search
provides an effective and generalizable framework for designing
accurate and compact VQE circuits in near-term quantum chemistry.
\end{abstract}

% ===================================================================
\section{Introduction}
% ===================================================================

The variational quantum eigensolver (VQE) has emerged as one of the
most promising near-term quantum algorithms for computing ground-state
energies of molecular
systems.\cite{Peruzzo2014,Tilly2022,Cerezo2021VQA,McArdle2020,Cao2019,Sajjan2022,Sheng2025Cavity}
At its core, VQE minimizes the expectation value of a molecular
Hamiltonian over a parameterized quantum state, the ansatz,
through a classical--quantum hybrid optimization loop.
The unitary coupled cluster singles and doubles (UCCSD) ansatz provides
a systematically improvable description of electron correlation and has
long served as the theoretical standard for quantum chemistry on quantum
devices.\cite{Bartlett2007,Anand2022,Aspuru-Guzik2005,Romero2019,Lee2019,Evangelista2019,Barkoutsos2018}
However, full UCCSD includes all single and double excitations from
occupied to virtual orbitals; for an $N$-orbital active space the
operator count scales as $\mathcal{O}(N^4)$, quickly rendering the
circuit too deep for current noisy intermediate-scale quantum (NISQ)
hardware.\cite{Wecker2015,Preskill2018,Kandala2017,O'Malley2016,Guo2024}

The adaptive derivative-assembled pseudo-Trotter ansatz VQE
(ADAPT-VQE)\cite{Grimsley2019} was introduced to address this depth
problem by constructing the ansatz iteratively: starting from the
Hartree--Fock state, it evaluates the energy gradient for each
candidate operator and appends the one with the largest gradient,
repeating until convergence. This greedy, sequential strategy produces
compact ans\"{a}tze that often outperform truncated UCCSD at the same
circuit depth, and has inspired a family of related methods including
qubit-ADAPT-VQE with reduced gradient evaluation
cost,\cite{Tang2021} qubit-excitation-based (QEB)
ADAPT,\cite{Yordanov2021} overlap-guided compact ansatz
selection,\cite{Feniou2023} and measurement-efficient variants with
improved parameter
optimizers.\cite{Li2024SOAP,Jiang2025SOAP}
Despite this progress, achieving both accuracy and compactness in UCC
ansatz design remains a combinatorially challenging problem: selecting
an optimal operator subset from the UCCSD pool involves navigating an
exponentially large discrete space, while simultaneously balancing
chemical accuracy against circuit depth and gate count.

Quantum architecture search (QAS) offers a complementary approach by
treating circuit design as a structured optimization problem over a
predefined operator pool, analogously to neural architecture search
(NAS) in classical machine
learning.\cite{Biamonte2017,Cerezo2022,Alexeev2025} Prior QAS studies
in a chemistry context have focused on hardware-efficient ans\"{a}tze
(HEA) for small systems of 2--6 qubits, with circuits comprising
generic single-qubit rotations and entangling gates without chemical
structure, using evolutionary, reinforcement-learning, or differentiable search
strategies.\cite{Du2022QAS,Wu2023QuantumDARTS,Ostaszewski2021,Wang2022QuantumNAS,Saib2023,Zhang2021NeuralQAS}
Other compact-ansatz approaches include machine learning-assisted
construction of shallow dynamic
ans\"{a}tze.\cite{Haider2024,Zeng2023,Li2025HQNN}
While these studies establish the feasibility of automated circuit
design, accuracy on chemically relevant systems remains well below
that of ADAPT-VQE and physics-constrained hardware-efficient
designs,\cite{Xiao2024,Kottmann2021} leaving open whether architecture search can match
or surpass greedy adaptive methods when applied to a structured
chemical operator pool.

Differentiable QAS (DQAS)\cite{Zhang2022DQAS,Liu2019,Chen2019} addresses the
discrete-selection bottleneck by relaxing operator choice into a
continuous optimization over architecture parameters, enabling joint
gradient-based optimization of circuit structure and variational
parameters in a single training loop. Applying DQAS to the UCCSD
operator pool reformulates UCC circuit design as a continuous
optimization problem.
By organizing operators into spin-paired groups that encode fermionic
symmetry and share variational parameters, the continuous relaxation
can jointly optimize all operator selections and circuit parameters
simultaneously, enabling the discovery of compact, globally optimized
UCC circuits that benefit from coordinated multi-operator optimization. We
introduce two complementary strategies: a
global mode that jointly optimizes all operator selections
simultaneously, and a layerwise mode that builds the circuit
incrementally while preserving previously learned structure.

In this work, we benchmark both strategies against ADAPT-VQE on
molecules spanning 8 to 14 qubits, including BeH$_2$, H$_4$, LiH, linear H$_6$, and
H$_2$O. DQAS achieves chemical accuracy at operator counts comparable to
ADAPT-VQE while producing circuits with fewer CNOT gates,
demonstrating that continuous relaxation of operator selection enables
more accurate and hardware-efficient ans\"{a}tze in the compact-circuit
regime. These results establish differentiable architecture search as
an effective and generalizable framework for UCC ansatz design on
near-term quantum devices.

% ===================================================================
\section{Theory and Methods}
% ===================================================================

\subsection{Variational Quantum Eigensolver}

VQE is a hybrid classical--quantum
algorithm designed to approximate the ground-state energy of a molecular
Hamiltonian $\hat{H}$ on near-term quantum hardware.\cite{Peruzzo2014,Tilly2022,Cerezo2021VQA,McArdle2020}
A parameterized quantum circuit $\hat{U}(\boldsymbol{\theta})$ acting on
a reference state $|\psi_0\rangle$ prepares the trial state
$|\psi(\boldsymbol{\theta})\rangle = \hat{U}(\boldsymbol{\theta})|\psi_0\rangle$.
The energy expectation value
\begin{equation}
  E(\boldsymbol{\theta}) = \langle\psi(\boldsymbol{\theta})\vert\hat{H}\vert\psi(\boldsymbol{\theta})\rangle
  \geq E_0
\end{equation}
provides an upper bound to the true ground-state energy $E_0$ for any
choice of $\boldsymbol{\theta}$, and is estimated on the quantum device
by decomposing $\hat{H}$ into a weighted sum of Pauli operators and
measuring each term separately. A classical optimizer iteratively
updates $\boldsymbol{\theta}$ to minimize $E(\boldsymbol{\theta})$,
with each iteration requiring one or more quantum circuit evaluations.

The performance of VQE depends critically on the structure of the
ansatz $\hat{U}(\boldsymbol{\theta})$. For quantum chemistry, the
UCCSD ansatz is a
chemically motivated choice in which the trial state is prepared by
applying products of parameterized exponentials of fermionic single and
double excitation operators to the Hartree--Fock reference state.
HEA, by contrast, use generic
single-qubit rotations and entangling gates arranged in a
hardware-adapted pattern, achieving shallow circuits without encoding
chemical structure. The DQAS framework developed in this work operates
as a meta-optimizer over the structure of the ansatz: it searches within
either the UCCSD operator pool or the HEA gate pool to identify the
subset of operations that minimizes the VQE energy at a target circuit
depth.

\subsection{UCCSD Operator Pool and Grouping}

We adopt the UCCSD operator pool as defined by TenCirChem,\cite{Li2023TenCirChem}
which generates single and double fermionic excitation operators from
an active space and maps them to qubits via the Jordan--Wigner
transformation.\cite{Jordan1928,Whitfield2011}
Each fermionic excitation operator is implemented using the efficient
circuit construction of Yordanov et al.,\cite{Yordanov2020} which
requires at minimum 2 and 8 CNOT gates for single and double excitations,
respectively, a linear reduction relative to the standard
CNOT-staircase construction. Alternative mappings
such as the Bravyi--Kitaev transformation\cite{Seeley2012} can reduce
qubit-operator weight in some settings; we adopt the Jordan--Wigner
convention throughout this work. For a given active space with
$N_e$ electrons in $N_o$ orbitals, the pool contains operators of the
form
\begin{equation}
  \hat{\tau}_\mu = \hat{a}^\dagger_p \hat{a}_q
    - \hat{a}^\dagger_q \hat{a}_p
  \quad \text{(singles)},
\end{equation}
\begin{equation}
  \hat{\tau}_\mu = \hat{a}^\dagger_p \hat{a}^\dagger_q \hat{a}_r \hat{a}_s
    - \hat{a}^\dagger_s \hat{a}^\dagger_r \hat{a}_q \hat{a}_p
  \quad \text{(doubles)},
\end{equation}
where $\hat{a}^\dagger$ and $\hat{a}$ are fermionic creation and
annihilation operators. In the compact tuple notation used throughout
this work, a single excitation is denoted $(p, q)$ and a double
excitation $(p, q, r, s)$, where $q$ (or $r, s$) are occupied and
$p$ (or $p, q$) are virtual spin-orbital indices. The spin-orbital
indices follow the TenCirChem convention: for $N_o$ active spatial
orbitals, $\alpha$ spin-orbitals carry indices $0,\ldots,N_o{-}1$ and
$\beta$ spin-orbitals carry indices $N_o,\ldots,2N_o{-}1$. A complete
description of the convention, including a decoded listing of all
operator groups for LiH, is provided in the Supporting Information
(Sections~S1--S2).

Because physical Hamiltonians preserve spin symmetry, each spatial
excitation naturally gives rise to a pair of spin-resolved operators
(spin-up and spin-down counterparts). We exploit this structure by
treating each such spin-pair as a single operator group. The
DQAS search therefore selects operator groups rather than individual
operators, halving the effective search space and enforcing spin
symmetry by construction. Each operator group contains either two
spin-paired operators (one $\alpha$-spin and one $\beta$-spin
excitation sharing a single variational parameter) or, for
spin-symmetric excitations whose $\alpha$- and
$\beta$-spin forms are identical by construction, a single operator.
Thus $k$ selected operator groups comprise between $k$ and $2k$
individual operators; all operator-group counts reported in this work
refer to groups, not individual operators. For the doubly exciting
operators, we further restrict the pool to operators generated from
the UCCSD excitation scheme with the \texttt{pick\_ex2} selection
of TenCirChem, which retains the most chemically relevant two-body
excitations.

As a fixed-ordering baseline, truncated UCCSD refers to a circuit
constructed by selecting the first $k$ operator groups from the full pool
in a predetermined canonical ordering, without any architecture
optimization; the variational parameters $\boldsymbol{\theta}$ are still
optimized by gradient-based optimization (BFGS) for a fair comparison.
When $k$ equals the total pool size, truncated UCCSD is identical to
full UCCSD; the operator-scaling plots presented in this work extend
to this limit, so that the full UCCSD circuit is included within the
range of the comparison.

\subsection{DQAS-Global for Ansatz Design}

\begin{figure}[htbp]
  \centering
  \includegraphics[width=\linewidth]{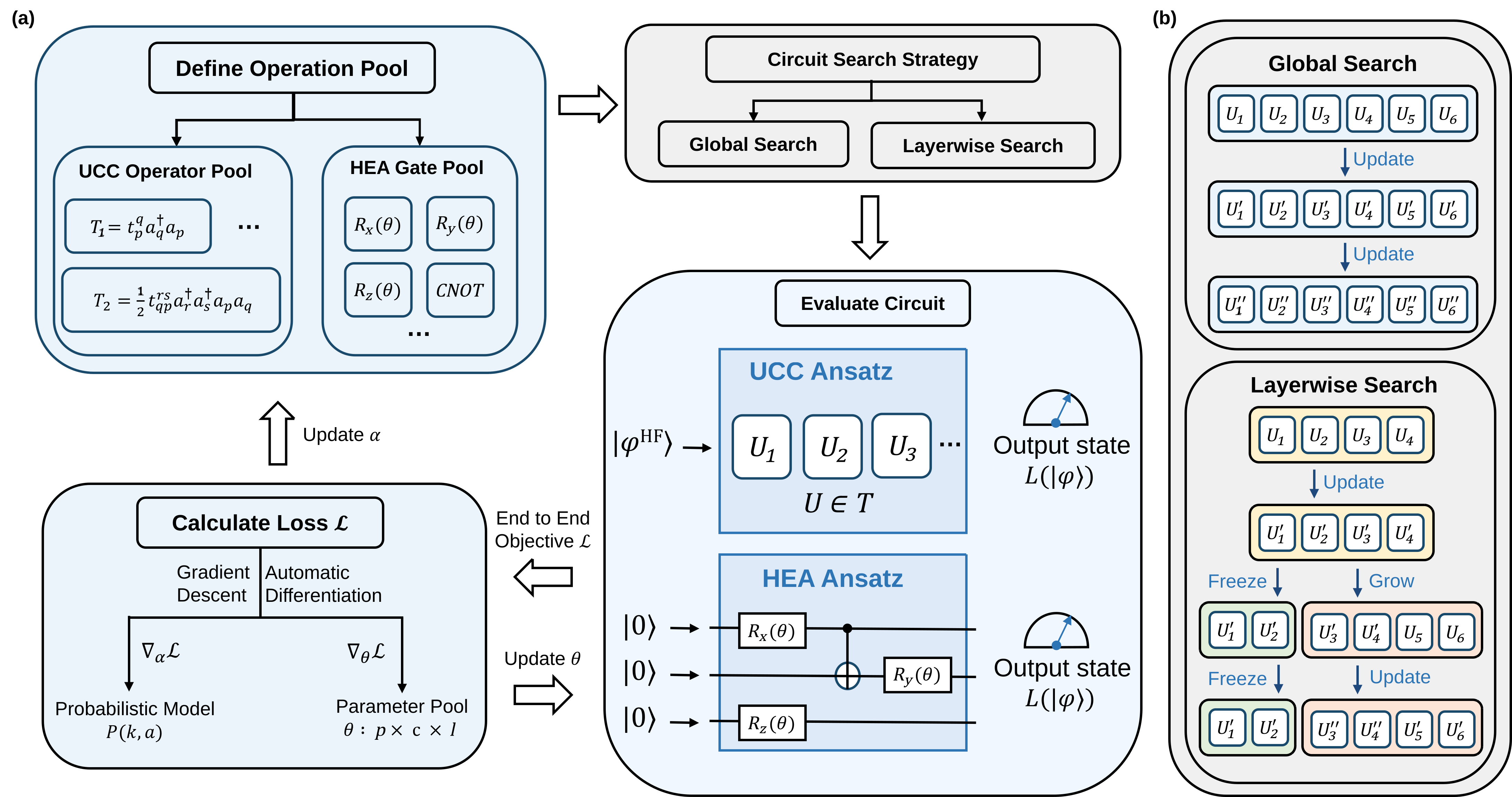}
  \caption{\textbf{Overview of the DQAS framework for quantum
  ansatz design.}
  (a) End-to-end DQAS workflow: (Step~1) operator pool definition,
  (Step~2) search strategy selection (global or layerwise),
  (Step~3) circuit evaluation and loss computation, and
  (Step~4) joint gradient update of architecture parameters
  $\alpha$ and variational parameters $\theta$.
  (b) Schematic comparison of global search, in which all circuit layers
  are optimized simultaneously, and layerwise search, in which the
  circuit grows incrementally with committed layers frozen and a sliding
  window of layers searched at each step.}
  \label{fig:fig1}
\end{figure}

In the global mode, we construct an $L$-layer ansatz where each layer
selects one operator group from the full pool of $M$ groups. Following
the DQAS formulation,\cite{Zhang2022DQAS} each layer $l$ maintains a
continuous architecture parameter vector $\boldsymbol{\alpha}^{(l)}
\in \mathbb{R}^M$, which defines a categorical probability distribution
over operator groups via a softmax:
\begin{equation}
  p(k_l = i \mid \boldsymbol{\alpha}^{(l)})
  = \mathrm{softmax}(\boldsymbol{\alpha}^{(l)})_i
  = \frac{e^{\alpha_i^{(l)}}}{\sum_{j=1}^{M} e^{\alpha_j^{(l)}}}.
  \label{eq:softmax}
\end{equation}
A circuit structure $\mathbf{k} = (k_1,\ldots,k_L)$ is sampled from
the factored distribution
\begin{equation}
  P(\mathbf{k} \mid \boldsymbol{\alpha})
  = \prod_{l=1}^{L} p(k_l \mid \boldsymbol{\alpha}^{(l)}),
  \label{eq:prob_model}
\end{equation}
and each sampled circuit prepares the state
$\hat{U}(\mathbf{k},\boldsymbol{\theta})\vert\psi_0\rangle
= \prod_{l=1}^{L} e^{\theta_{k_l}\hat{\tau}_{k_l}}\vert\psi_0\rangle$.
The joint optimization objective is the expected energy over sampled
circuits:
\begin{equation}
  \mathcal{L}(\boldsymbol{\theta},\boldsymbol{\alpha})
  = \mathbb{E}_{\mathbf{k}\sim P(\mathbf{k}|\boldsymbol{\alpha})}
    \bigl[E(\mathbf{k},\boldsymbol{\theta})\bigr],
  \quad
  E(\mathbf{k},\boldsymbol{\theta})
  = \langle\psi_0\vert
      \hat{U}^{\dagger}(\mathbf{k},\boldsymbol{\theta})\,\hat{H}\,
      \hat{U}(\mathbf{k},\boldsymbol{\theta})
    \vert\psi_0\rangle.
  \label{eq:objective}
\end{equation}
At each training step, a batch of $K$ circuit structures is sampled
from $P(\mathbf{k}\mid\boldsymbol{\alpha})$; the gradient with respect
to $\boldsymbol{\theta}$ is obtained by reverse-mode automatic
differentiation through the sampled circuits, while the gradient with
respect to $\boldsymbol{\alpha}$ is estimated by the naive mean-field
(NMF) score-function estimator\cite{Zhang2022DQAS} with a batch
baseline to reduce variance. Both parameter sets are updated jointly
with the Adam optimizer.\cite{Kingma2015Adam}

After the search converges, a discrete ansatz is extracted by
selecting the operator group with the highest architecture parameter
at each layer:
\begin{equation}
  i^{(l)}_\mathrm{select} = \underset{i}{\arg\max}\;
  \alpha_i^{(l)}.
  \label{eq:argmax}
\end{equation}
The discrete circuit is then fine-tuned by re-optimizing the
variational parameters $\boldsymbol{\theta}$ with the architecture
fixed (BFGS). This two-stage procedure, probabilistic search followed by
discrete fine-tuning, is standard in differentiable architecture
search and ensures that the final circuit corresponds to a
physically realizable gate sequence.

To mitigate the risk of local optima, the search-and-fine-tune
procedure is repeated over multiple independent initializations;
hyperparameter details are provided in Section~S4 of the Supporting
Information.
While effective for moderate circuit depths, the global approach
requires simultaneous optimization of all $L$ layers, which becomes
computationally demanding as $L$ grows; the layerwise strategy
introduced next addresses this limitation by decomposing the search
into a sequence of smaller, tractable subproblems.

\subsection{DQAS-Layerwise for Ansatz Design}

The global mode treats the full $L$-layer ansatz as a single
optimization problem, which can become computationally demanding as
the number of layers grows. The layerwise mode addresses this by
constructing the circuit incrementally via a sliding-window strategy.
At each growth step, a window of $k$ soft layers is searched jointly:
the first $|\mathrm{warm}|$ layers are warm-started from the carry-over
buffer of the previous step, and the remaining layers are freshly
initialized with low-amplitude Gaussian noise
$\boldsymbol{\alpha} \sim \mathcal{N}(0,\sigma^2)$.
After the search, only $s < k$ new layers are committed (frozen) to
the circuit, while the remaining $k - s$ layers are retained as the
warm-start buffer for the next step; their fine-tuned variational
parameters are also carried over.
Following each commitment, the variational parameters of the entire
circuit are re-optimized with the architecture fixed using BFGS,
and the frozen parameter set $\boldsymbol{\theta}_\mathrm{fixed}$ is updated.
The full procedure is given in Algorithm~\ref{alg:layerwise}.

\begin{algorithm}[htbp]
\caption{DQAS-Layerwise for UCCSD Ansatz Design}
\label{alg:layerwise}
\begin{algorithmic}[1]
\Require Operator pool $\mathcal{P} = \{\hat{\tau}_1,\ldots,\hat{\tau}_M\}$,
         target depth $L$, window size $k$, slide amount $s$ ($s < k$), epochs $T$
\Ensure  Discrete circuit $\mathcal{C}$ with $L$ operator groups
\State $\mathcal{C} \leftarrow \emptyset$,\;
       $\boldsymbol{\theta}_\mathrm{fixed} \leftarrow \emptyset$,\;
       $\mathrm{warm} \leftarrow \emptyset$ \Comment{circuit; frozen params; warm-start carry-over buffer}
\While{$|\mathcal{C}| < L$}
  \State \textbf{// --- Search phase ---}
  \State Initialize architecture params
         $\boldsymbol{\alpha}^{(1)},\ldots,\boldsymbol{\alpha}^{(k)}
         \sim \mathcal{N}(0,\sigma^2)$,
         warm-starting layers $1,\ldots,|\mathrm{warm}|$
         from $\mathrm{warm}$;
         initialize variational params
         $\boldsymbol{\theta}^{(1)},\ldots,\boldsymbol{\theta}^{(k)}$
  \For{$t = 1$ \textbf{to} $T$}
    \State Sample batch of structures
           $\mathbf{k}^{(1)},\ldots,\mathbf{k}^{(K)}
           \sim \prod_{j=1}^{k} p(k_j \mid \boldsymbol{\alpha}^{(j)})$
           using Eq.~\protect\eqref{eq:prob_model}
    \State Compute batch average energy
           $\bar{E} = \frac{1}{K}\sum_{b=1}^{K}
           E(\mathcal{C} \cup \mathbf{k}^{(b)},
           \boldsymbol{\theta}_\mathrm{fixed}, \boldsymbol{\theta}^{(1:k)})$
           using Eq.~\protect\eqref{eq:objective}
    \State Update $\{\boldsymbol{\alpha}^{(j)},\boldsymbol{\theta}^{(j)}\}_{j=1}^k$
           via Adam, keeping $\boldsymbol{\theta}_\mathrm{fixed}$ frozen
  \EndFor
  \State \textbf{// --- Commit $s$ new layers ---}
  \For{$j = 1$ \textbf{to} $s$}
    \State $i^\star \leftarrow \arg\max_i\;
           \alpha_i^{(j)}$
    \State Append hard layer $e^{\theta_{i^\star}\hat{\tau}_{i^\star}}$
           to $\mathcal{C}$
  \EndFor
  \State \textbf{// --- Fine-tune ---}
  \State Re-optimize all variational params of $\mathcal{C}$ with
         architecture fixed (BFGS); update $\boldsymbol{\theta}_\mathrm{fixed}$
  \State \textbf{// --- Update carry-over buffer ---}
  \State $\mathrm{warm} \leftarrow$ last $k - s$ entries of $\mathcal{C}$
         with their fine-tuned $\boldsymbol{\theta}$ values
\EndWhile
\State \Return $\mathcal{C}$
\end{algorithmic}
\end{algorithm}

By freezing previously learned layers during each new search phase,
the layerwise approach preserves earlier structural choices and focuses
the optimization capacity on newly added operators. This progressive
construction, drawing on layerwise learning strategies developed for
quantum neural networks\cite{Skolik2021} and progressive differentiable
NAS,\cite{Chen2019} reduces the effective dimensionality of each search
step
and generally finds more accurate ans\"{a}tze than the global mode
for larger, more correlated systems such as H$_6$ and H$_2$O. The
grouping strategy introduced in Section~2.2 is retained in the
layerwise mode: each selected layer adds a pair of spin-matched
operators, maintaining physical symmetry throughout the construction.
As with the global mode, the layerwise procedure is repeated over
multiple independent initializations per growth step; hyperparameter
details are provided in Section~S4 of the Supporting Information.

% ===================================================================
\section{Computational Details}
% ===================================================================

All molecular Hamiltonians were computed at the Hartree--Fock level
using PySCF\cite{Sun2020PySCF} with the STO-3G basis set. The
qubit Hamiltonian was obtained via the Jordan--Wigner transformation
as implemented in OpenFermion\cite{McClean2020OpenFermion} and
TenCirChem.\cite{Li2023TenCirChem} The full configuration interaction
(FCI) reference energies were computed by exact diagonalization of the
sparse qubit Hamiltonian. All quantum circuit simulations were
performed using TensorCircuit\cite{Zhang2023TC,Zhang2025TCNG} with the JAX backend
in \texttt{complex128} precision.

The molecular systems studied in this work are listed in
Table~\ref{tab:systems}. Architecture and variational parameters were
jointly optimized using the Adam optimizer\cite{Kingma2015Adam} during
the search phase; discrete ansatz parameters were subsequently
fine-tuned using BFGS.
Full hyperparameter and initialization details are provided in
Section~S4 of the Supporting Information. The CNOT count for each operator group is
computed from the Jordan--Wigner decomposition of the corresponding
UCC exponential.\cite{Yordanov2020} For the truncated UCCSD baseline, operator groups are
ordered by decreasing second-order M\o{}ller--Plesset perturbation theory (MP2) amplitude magnitude, and the first $k$ groups
are selected. A subset of H$_2$O calculations was repeated using
the qubit-excitation-based (QEB) operator pool\cite{Tang2021,Yordanov2021} to assess
the benefit of a hardware-aware operator pool. HEA benchmarks were
performed on H$_2$ (4 qubits, JW) and LiH under parity (4 qubits)
and Jordan--Wigner (6 qubits) encodings; HEA gate pool and
circuit details are provided in Section~S4 of the Supporting
Information.

\begin{table}[htbp]
\centering
\caption{Molecular systems studied in this work (UCCSD and QEB pool benchmarks).
Active spaces use the notation ($N_e$ electrons, $N_o$ orbitals); qubit counts
assume the Jordan--Wigner encoding unless otherwise noted. Hardware-efficient
ansatz (HEA) benchmarks are listed in Table~S2 of the Supporting Information.}
\label{tab:systems}
\begin{tabular}{llcl}
\toprule
\textbf{Molecule} & \textbf{Active space} & \textbf{Qubits} & \textbf{Method} \\
\midrule
H$_4$ (linear)  & (4e, 4o) &  8 & DQAS-Global \\
BeH$_2$         & (4e, 5o) & 10 & DQAS-Global \\
LiH             & (4e, 6o) & 12 & DQAS-Global \\
BeH$_2$         & (4e, 6o) & 12 & DQAS-Layerwise \\
H$_6$ (linear)  & (6e, 6o) & 12 & DQAS-Layerwise \\
H$_2$O          & (4e, 8o) & 14 & DQAS-Layerwise, QEB \\
\bottomrule
\end{tabular}
\end{table}

% ===================================================================
\section{Results and Discussion}
% ===================================================================

\subsection{DQAS Search Dynamics}

\begin{figure}[htbp]
  \centering
  \includegraphics[width=\linewidth]{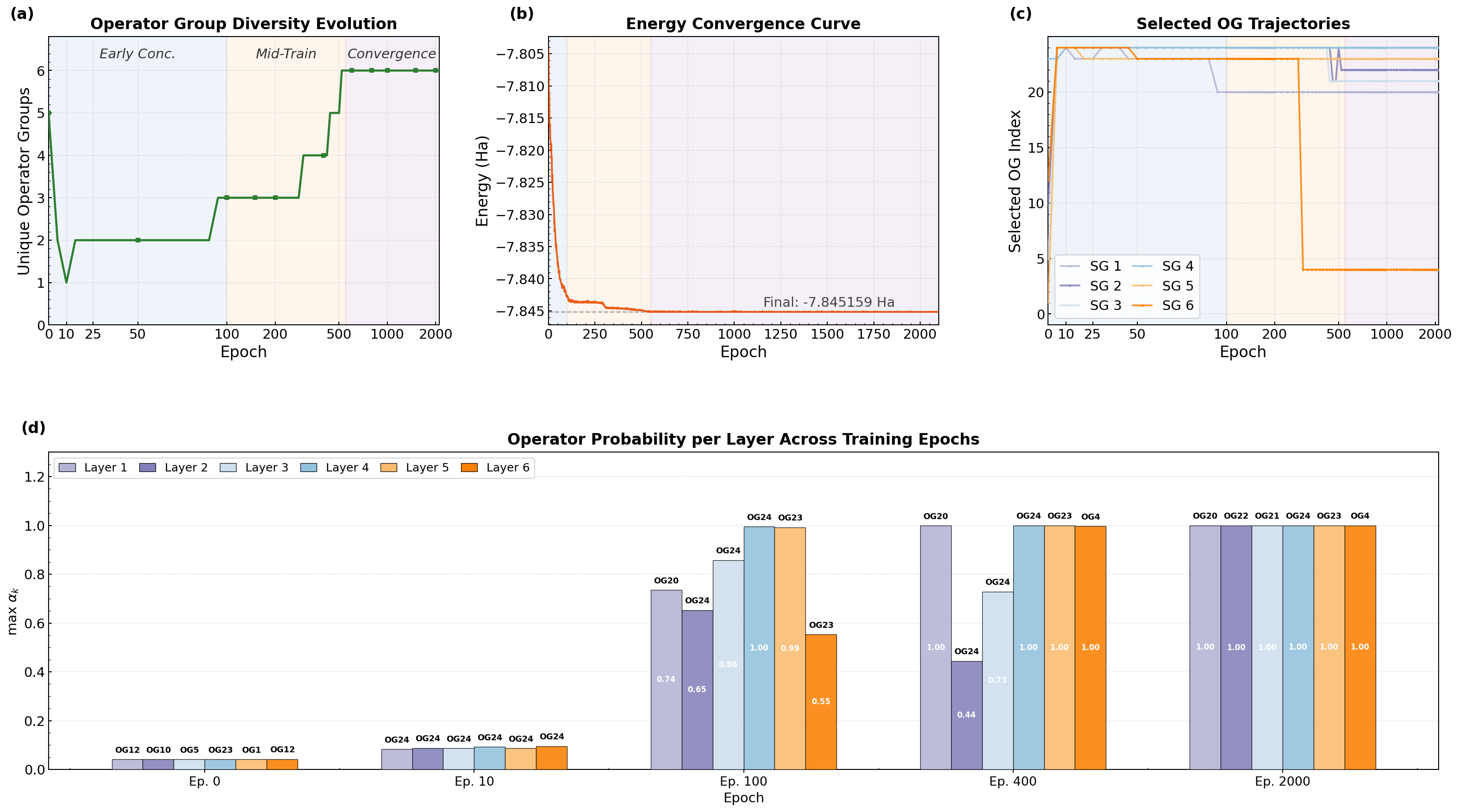}
  \caption{\textbf{DQAS-Global search dynamics for LiH ($d = 2.20$ \AA,
  6-operator-group ansatz).}
  (a) Operator group (OG) diversity, measured as the number of distinct OGs sampled per
  batch, versus training epoch.
  (b) Variational energy convergence toward the FCI reference
  ($-7.8454$ Ha, gray dashed line).
  (c) Selected OG trajectories for layers SG\,1--SG\,6.
  (d) Maximum architecture-parameter probability ($\max_k\,\alpha_k$)
  for each of the six layers at five representative training epochs
  (Ep.\,0, 10, 100, 400, 2000), grouped by epoch; bar labels indicate
  the dominant operator group index.}
  \label{fig:fig2}
\end{figure}

Before analyzing energetic performance, we first examine the internal
search dynamics of DQAS-Global using LiH as an illustrative example,
focusing on the typical training trajectory and convergence behavior
of the algorithm. Figure~\ref{fig:fig2}
presents the training trajectory for LiH at a stretched geometry
($d = 2.20$ \AA) with a 6-operator-group ansatz. The operator group (OG)
index axis in panel~(c) refers to the 25-group pool enumerated
in Table~S1 of the Supporting Information; at convergence, the six
selected groups are \{4, 20, 21, 22, 23, 24\}.

Three distinct phases emerge consistently across all training runs.
In the early concentration phase (the first $\sim$50 epochs), the
nearly uniform initial probability distribution, arising from
low-amplitude Gaussian noise initialization of the architecture
parameters $\boldsymbol{\alpha}$, rapidly collapses onto the small
subset of operator groups whose energy gradients are largest. This
mirrors the behavior of greedy methods such as ADAPT-VQE but occurs
simultaneously across all layers. In the mid-training phase (epochs 50--500), the probabilistic model recovers diversity
as gradients with respect to $\boldsymbol{\alpha}$ drive exploration
of alternative operator combinations; the number of unique groups
sampled per batch increases and the energy continues to decrease.
In the final convergence phase (epochs $>$500), each layer
locks onto a distinct, high-probability operator group, and the energy
plateaus near the FCI value of $-7.8454$ Ha. The selected-OG
trajectories in panel (c) confirm that layer assignments are stable
by epoch 1000, and panel~(d) displays the maximum
architecture-parameter probability for each layer at five
representative epochs, confirming that by epoch 2000 every layer
has converged to a single dominant operator group with probability
$>0.99$.

This three-stage dynamics, comprising early concentration, diversity recovery, and stable
convergence, is an emergent
property of the mean-field probabilistic model and is reproduced across
independent runs, confirming that DQAS training is both effective and
robust. The diversity recovery phase is particularly important, as it highlights
a key advantage of DQAS: by simultaneously exploring operator
assignments across all layers, the algorithm identifies better circuit
structures beyond what single-step gradient selection would yield.

\subsection{DQAS-Global: Potential Energy Surfaces and Operator Scaling}

Having characterized the search dynamics of DQAS-Global, we now evaluate
energetic performance and operator efficiency across three molecules.

\begin{figure}[htbp]
  \centering
  \includegraphics[width=\linewidth]{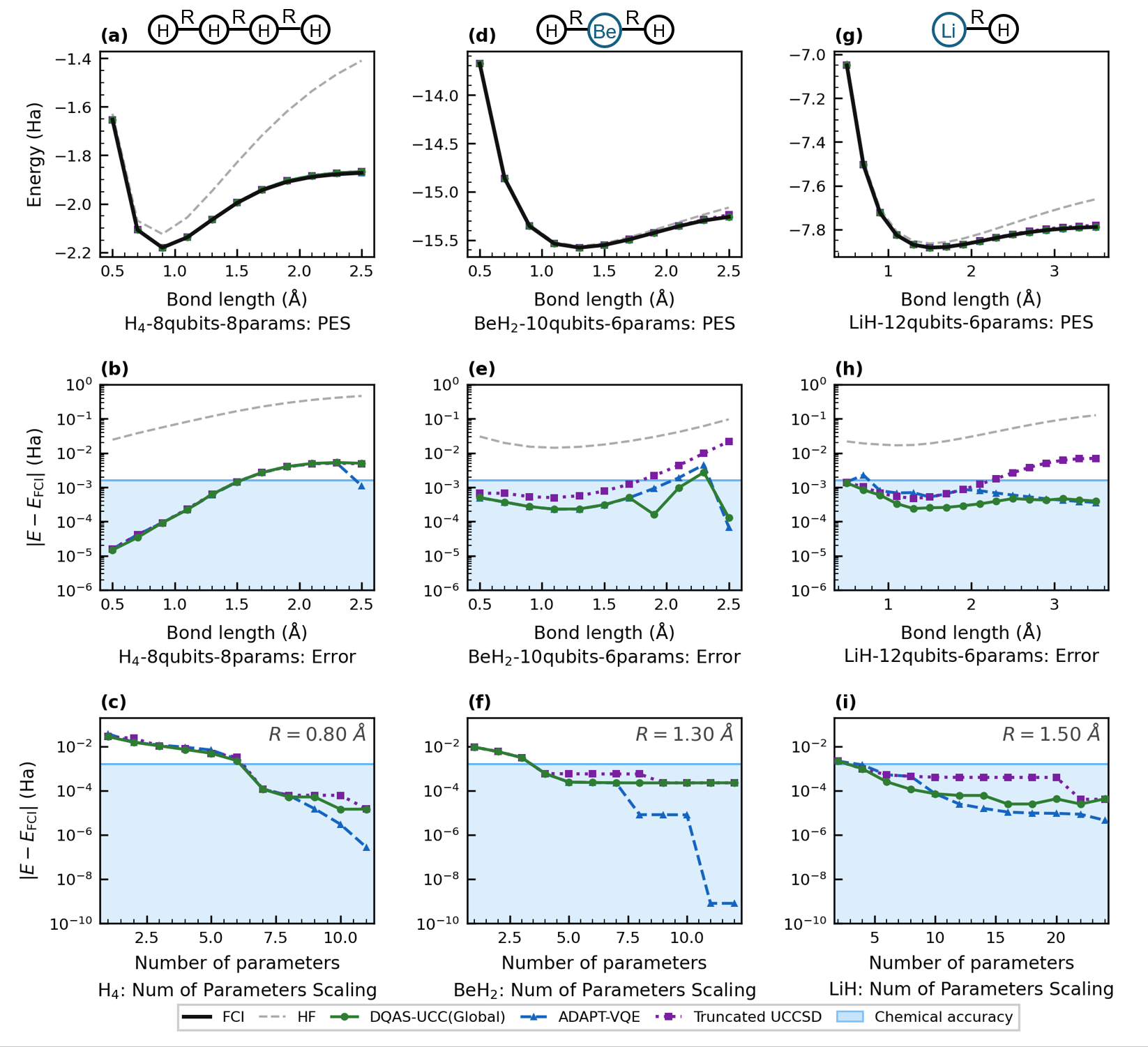}
  \caption{\textbf{DQAS-Global performance on H$_4$ (8 qubits),
  BeH$_2$ (10 qubits), and LiH (12 qubits).}
  Columns correspond to H$_4$, BeH$_2$, and LiH, respectively.
  (a,d,g) Potential energy surfaces.
  (b,e,h) Absolute energy error relative to FCI versus bond length.
  (c,f,i) Energy error as a function of operator group count at a
  representative near-equilibrium geometry.
  The blue shaded band marks chemical accuracy (1.6 mHa).}
  \label{fig:fig3}
\end{figure}

Figure~\ref{fig:fig3} shows the potential energy curves (PECs) and
operator-scaling results obtained by DQAS-Global for BeH$_2$, H$_4$,
and LiH, alongside ADAPT-VQE and truncated UCCSD references.
A key advantage of DQAS-Global over both ADAPT-VQE and na\"{i}vely
truncated UCCSD is its ability to learn which operators matter
most for the specific molecular geometry, rather than selecting
by instantaneous gradient or fixed ordering. For BeH$_2$ at 6 operator
groups ($d = 1.30$ \AA), truncated UCCSD achieves an error of
0.57 mHartree (mHa), more than twice the DQAS-Global error, while
ADAPT-VQE also lags DQAS-Global at intermediate circuit depths.
The operator-scaling results below illustrate this advantage
quantitatively across all three molecules.

The PECs for all three molecules demonstrate that DQAS-Global closely
tracks the FCI reference, achieving chemical accuracy at most bond
lengths with a fixed operator group count.
For H$_4$ (a linear hydrogen chain frequently used
as a benchmark for strong correlation), the DQAS-Global energy closely
follows ADAPT-VQE from compressed to stretched geometries, with both
methods capturing the onset of static correlation that characterizes the
dissociation regime. The operator-scaling behavior, shown in the bottom
row of Figure~\ref{fig:fig3}, plots the energy error relative to FCI as
a function of operator group count at a representative geometry for each
molecule.

At the representative geometry $d = 0.80$ \AA\ for H$_4$ (linear chain), DQAS-Global exhibits a
clear advantage over ADAPT-VQE at intermediate circuit depths. At 4
operator groups, DQAS-Global achieves an energy error of 7.26 mHa
versus 9.25 mHa for ADAPT-VQE, a 22\% reduction. This advantage
widens to 29\% at 5 groups (DQAS: 4.91 mHa; ADAPT-VQE:
6.93 mHa), and both methods converge to the same solution at 7
groups (error $\approx 0.12$ mHa, well within chemical accuracy).
ADAPT-VQE appends operators one at a time guided by the instantaneous
energy gradient, which tends to prioritize operators whose individual
contribution to the current energy is largest, without accounting for
combinations whose collective benefit exceeds their individual gradients.
DQAS-Global, by contrast, jointly optimizes all architectural
parameters from the outset, enabling the search to discover globally
complementary operator combinations.

For LiH ($d = 1.50$ \AA), DQAS-Global at 6 operator groups yields an
error of 0.25 mHa versus 0.52 mHa for ADAPT-VQE, a
two-fold improvement, while both methods satisfy chemical accuracy at
6 groups. For BeH$_2$ ($d = 1.30$ \AA), DQAS-Global and ADAPT-VQE
converge to nearly identical accuracy at all operator counts, with both
achieving errors well within chemical accuracy at 4 groups
(error $< 0.60$ mHa).
This molecule-dependent behavior reflects the degree of correlation
complexity: BeH$_2$ at equilibrium is well-described by a small number
of dominant excitations accessible to both strategies, whereas H$_4$
and LiH involve richer correlation structure that rewards a global
rather than greedy search.

We also note that at larger operator counts, DQAS-Global can fall
slightly below ADAPT-VQE in accuracy. As the number of selected groups
grows, the circuit sampled during DQAS training becomes progressively deeper, plausibly increasing susceptibility to
vanishing gradients\cite{McClean2018Barren,Arrasmith2022} with respect to both the architecture parameters
$\boldsymbol{\alpha}$ and the variational parameters $\boldsymbol{\theta}$,
though we have not directly measured gradient norms in this work.
ADAPT-VQE is less susceptible because at each addition step the
gradient is evaluated on a fully optimized current state, providing
an unambiguous selection signal regardless of circuit depth. The
practical implication is that DQAS-Global is most advantageous at
compact circuit depths, where its joint optimization strategy yields
circuits that benefit from simultaneous optimization of all operator
positions, precisely the
regime that aligns with the primary goal of obtaining compact,
hardware-efficient ans\"{a}tze for near-term quantum devices.

\subsection{DQAS-Layerwise: Enhanced Convergence for Larger Molecules}

\begin{figure}[htbp]
  \centering
  \includegraphics[width=\linewidth]{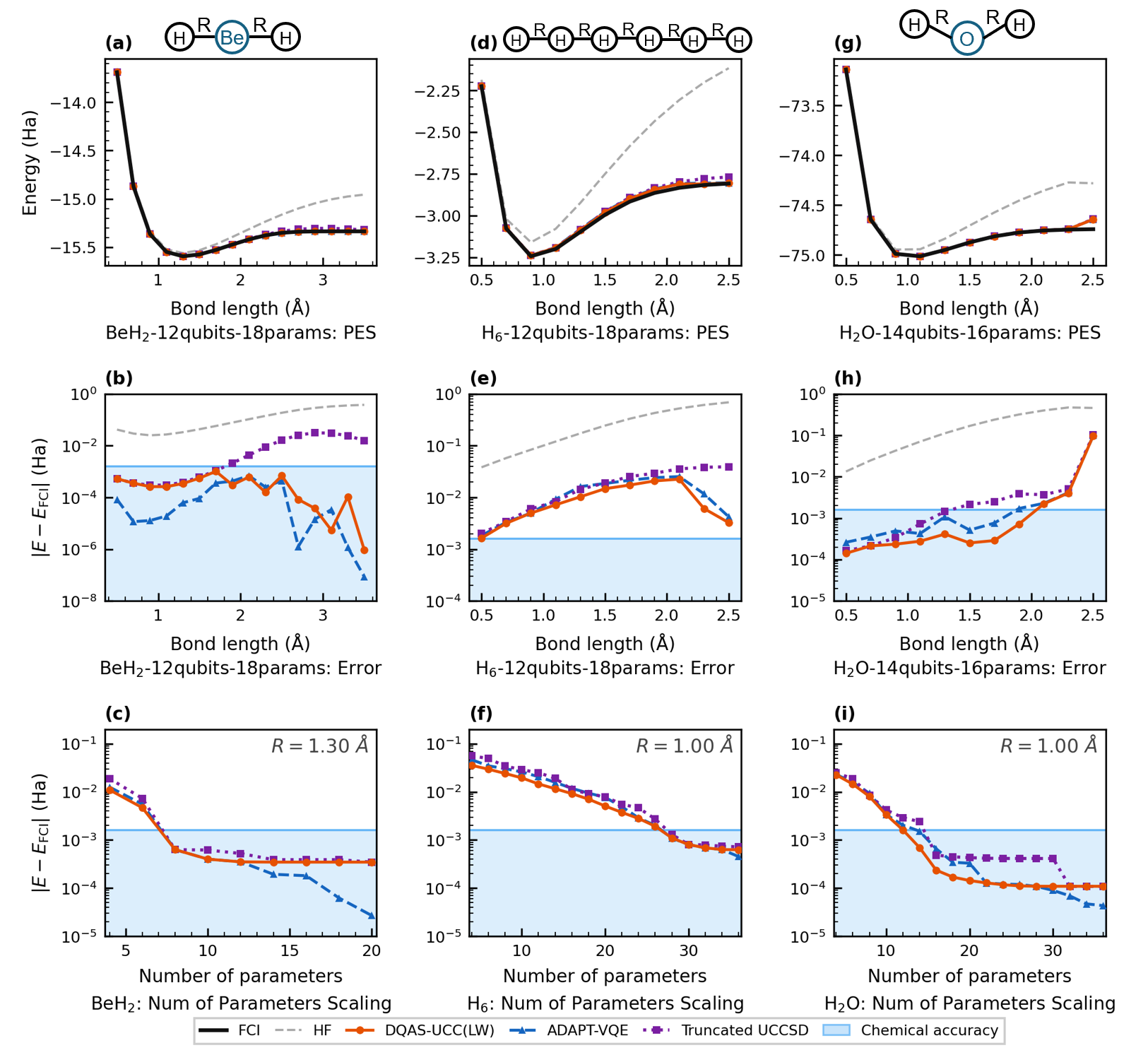}
  \caption{\textbf{DQAS-Layerwise performance on BeH$_2$ (12 qubits),
  H$_6$ (12 qubits), and H$_2$O (14 qubits).}
  Columns correspond to BeH$_2$, H$_6$, and H$_2$O, respectively.
  (a,d,g) Potential energy surfaces.
  (b,e,h) Absolute energy error relative to FCI versus bond length.
  (c,f,i) Energy error as a function of operator group count at a
  representative near-equilibrium geometry.
  The blue shaded band marks chemical accuracy (1.6 mHa).}
  \label{fig:fig4}
\end{figure}

These results indicate that DQAS-Global generally achieves higher
accuracy at equivalent compact operator counts compared to ADAPT-VQE
for small molecules; we next examine whether this advantage extends to
larger, more correlated systems using the layerwise strategy.
Figure~\ref{fig:fig4} presents the PECs and operator-scaling profiles
for DQAS-Layerwise applied to BeH$_2$ (expanded active space, 12
qubits), H$_6$ (12 qubits), and H$_2$O (14 qubits), compared against
ADAPT-VQE and truncated UCCSD.

For BeH$_2$ with the expanded active space (12 qubits), both DQAS-Layerwise and ADAPT-VQE reach chemical accuracy simultaneously at
8 operator groups, each achieving an energy error of 0.63 mHa
at $d = 1.3$ \AA. Beyond this point, ADAPT-VQE continues to improve more
rapidly with circuit depth, reaching 0.19 mHa at 14 operator
groups compared to 0.35 mHa for DQAS-Layerwise, consistent with
the gradient attenuation effect discussed above. The PEC
(Figure~\ref{fig:fig4}, column 1, top) confirms that both methods
track the FCI reference closely throughout the dissociation profile,
with maximum deviations below 0.7 mHa near equilibrium.

The most compelling advantage of DQAS-Layerwise emerges for water
(H$_2$O, 14 qubits). At 14 operator groups, both methods satisfy
the chemical accuracy threshold; however, DQAS-Layerwise achieves
an energy error of 0.70 mHa, more than twice as accurate as
ADAPT-VQE (1.52 mHa), at the same operator count. At 16 operator
groups, this advantage widens further: DQAS-Layerwise reaches
0.24 mHa versus 0.65 mHa for ADAPT-VQE, a 2.7-fold
improvement. The accuracy gap reflects the advantage of the DQAS search strategy:
by exploring operator combinations rather than selecting operators one
at a time, DQAS captures latent combinatorial correlations between
operators that are not apparent from single-operator gradient
evaluation.

For H$_6$ (linear hydrogen chain, 6 electrons, 12 qubits), a
challenging benchmark due to strong static correlation along the
dissociation coordinate, DQAS-Layerwise achieves an energy error of
1.64 mHa near the compressed geometry ($d = 0.5$ \AA), at the
boundary of chemical accuracy. Both DQAS-Layerwise and ADAPT-VQE
show increasing deviations from FCI as the bond stretches,
a behavior inherent to single-reference coupled-cluster methods in the
strongly correlated limit,\cite{Sokolov2020} with energy errors of 3.17 mHa versus
3.32 mHa for ADAPT-VQE at $d = 0.7$ \AA. The operator-scaling
panel (Figure~\ref{fig:fig4}, column 2, bottom) shows that
DQAS-Layerwise achieves lower energy errors than ADAPT-VQE at most
operator counts examined: at 18 operator groups the respective errors are
7.07 mHa and 9.44 mHa. While chemical accuracy for H$_6$
requires deeper circuits than those examined here, a consequence of
the strong static correlation inherent to this system; extended
calculations confirm that both DQAS-Layerwise and ADAPT-VQE reach
chemical accuracy at approximately 28 operator groups (out of the
39 groups in the full H$_6$ UCCSD pool). The primary goal of the
benchmarks shown in Figure~\ref{fig:fig4} is to characterize
performance in the compact-circuit regime rather than to pursue the
deepest possible circuits.
The general advantage of DQAS-Layerwise at compact operator counts
suggests that the global search strategy tends to deliver higher
accuracy per operator group, particularly in moderately and strongly
correlated regimes. For H$_6$, the truncated UCCSD
baseline performs similarly to ADAPT-VQE across most operator counts
(9.12 mHa versus 9.44 mHa at 18 groups), with both
substantially less accurate than DQAS-Layerwise, reflecting the
difficulty of selecting the most important operators under strong
correlation by either greedy or fixed-order strategies.

\subsection{Circuit Efficiency: Structural Origin of the CNOT Advantage}
\label{sec:cnot}

\begin{figure}[htbp]
  \centering
  \includegraphics[width=\linewidth]{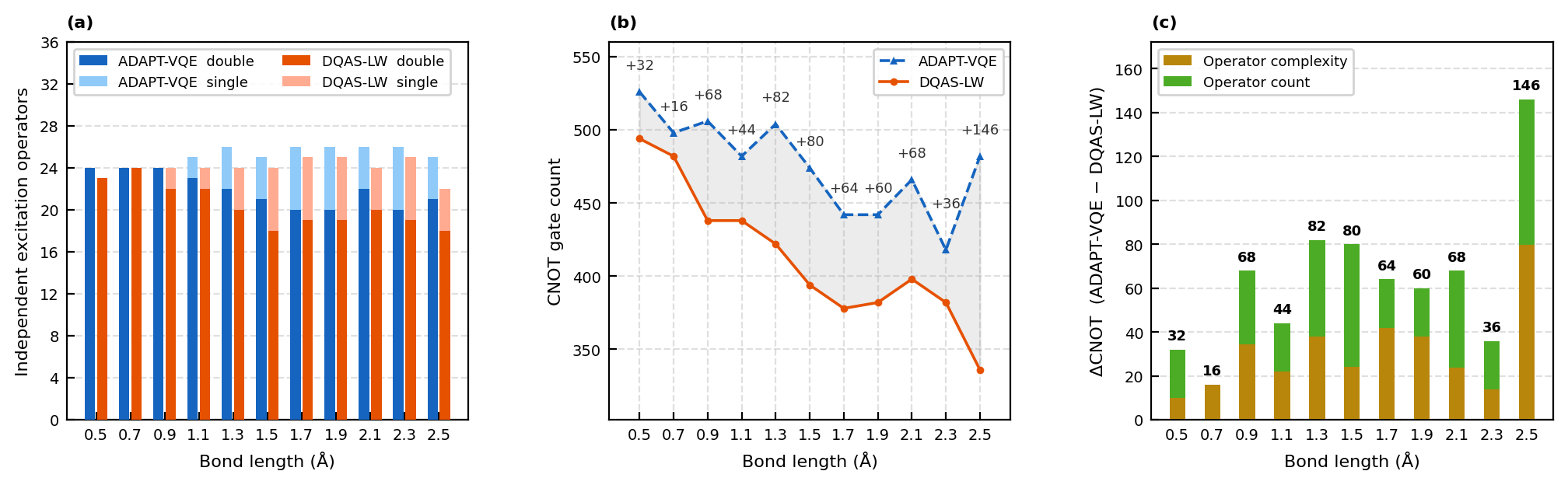}
  \caption{\textbf{Circuit composition analysis for H$_2$O (UCCSD pool,
  16 operator groups): DQAS-Layerwise versus ADAPT-VQE.}
  (a) Number of independent single and double excitation operators
  selected by each method versus bond length.
  (b) CNOT gate count versus bond length; annotations show
  $\Delta$CNOT (ADAPT-VQE $-$ DQAS-Layerwise) at each geometry.
  (c) Decomposition of $\Delta$CNOT into an operator-count contribution
  (green, arising from differences in the number and type of selected
  operators) and an operator-complexity contribution (tan, arising from
  differences in the qubit span of the selected operators).}
  \label{fig:fig5}
\end{figure}

Having established that DQAS-Layerwise achieves higher accuracy than
ADAPT-VQE at equivalent operator counts, we now examine the structural
basis for its simultaneous advantage in circuit gate count.
Figure~\ref{fig:fig5} presents a three-panel circuit-composition
analysis for H$_2$O (14 qubits, UCCSD pool, 16 operator groups) across
the full dissociation profile.

Panel~(a) reveals a consistent compositional difference: DQAS-Layerwise
selects systematically more single-excitation operators relative to
ADAPT-VQE across most geometries. Near equilibrium ($d = 1.5$ \AA),
DQAS-Layerwise selects 6 singles and 18 doubles (24 operators total),
while ADAPT-VQE selects 4 singles and 21 doubles (25 operators).
Since each spin-paired operator group comprises two constituent
operators (an $\alpha\alpha$ and a $\beta\beta$ excitation), the CNOT
analysis decomposes each selected group into its individual operators
before computing gate counts; the 24 operators at $d = 1.5$ \AA\
therefore correspond to 6 singles and 18 doubles at the
individual-operator level.
Crucially, the two methods share 21 of these operators; the entire
CNOT difference originates from the 3 operators unique to
DQAS-Layerwise and the 4 operators unique to ADAPT-VQE at this
geometry.
Since single excitations require only 2--10 CNOT gates under the
Jordan--Wigner decomposition,\cite{Yordanov2020} compared to 14 or
more for double excitations, this operator-type bias directly reduces
the total gate count.

Panel~(b) shows the resulting CNOT counts across the PES:
DQAS-Layerwise requires fewer CNOT gates than ADAPT-VQE at every
geometry, with the advantage ($\Delta$CNOT) ranging from 16 at
$d = 0.7$ \AA\ to 146 at $d = 2.5$ \AA.
Specifically, near equilibrium ($d = 0.9$--$1.5$ \AA), DQAS-Layerwise uses
394--438 CNOT gates versus 474--506 for ADAPT-VQE, reductions of
13--17\%. The gap widens markedly in the stretched regime, where the
two methods diverge most strongly in their operator selections as the
correlation structure changes.

Panel~(c) decomposes $\Delta$CNOT into two additive contributions,
illustrated at $d = 1.5$ \AA\ where $\Delta$CNOT = 80.
The operator-count contribution (green) captures the gate
savings from differences in the number and excitation type of the
selected operators, estimated using the pool-averaged CNOT costs
(5.20 CNOTs per single excitation, 22.05 CNOTs per double excitation);
this accounts for approximately 56 of the 80 CNOT difference.
The operator-complexity contribution (tan) captures the residual
savings ($\approx$24 CNOTs) attributable to differences in qubit span:
the 4 operators unique to ADAPT-VQE are all long-range double
excitations from the lowest-energy core orbitals (qubits 0--1) to the
highest virtual orbitals (qubits 12--13), spanning 11--13 qubits and
incurring 26--30 CNOTs each (total: 112 CNOTs). By contrast, the 3
operators unique to DQAS-Layerwise consist of two short-range single
excitations (qubit span $\approx$5, 6 CNOTs each) and one
medium-range double excitation (qubit span $\approx$8, 20 CNOTs),
totaling only 32 CNOTs, an 80-CNOT difference that matches the
observed $\Delta$CNOT exactly.

The physical origin of this selectivity lies in the difference between
joint and sequential operator selection.
DQAS-Layerwise optimizes all circuit positions simultaneously,
searching over combinations of operators rather than selecting each
greedily against a fixed context. Within this joint search, cheap
short-range excitations compete directly with expensive long-range
ones; the former can collectively provide equivalent correlation
energy through complementary contributions that are invisible to
single-operator gradient evaluation, giving them a systematic
advantage over individually more expensive alternatives.
ADAPT-VQE's sequential criterion, by contrast, evaluates each operator
against the current partial circuit without accounting for such
combinatorial trade-offs, and may therefore select operators whose
gate cost could be matched by cheaper alternatives in a jointly
optimized context.
This advantage may not generalize to strongly
correlated systems where specific long-range excitations make
essential and irreplaceable contributions; in such cases ADAPT-VQE's
sequential criterion may be better suited to identifying the critical
operators.

\subsection{CNOT-Efficient Ansatz Exploration}
\label{sec:qop}

Beyond UCCSD, two classes of gate-efficient circuits offer further
reductions in CNOT overhead: the qubit-excitation-based (QEB) operator
pool\cite{Tang2021,Yordanov2021}, which retains the chemically motivated
excitation structure of UCC while requiring substantially fewer CNOT
gates per operator, and hardware-efficient ans\"{a}tze (HEA), which
replace chemistry-specific operators entirely with hardware-native gate
primitives. Applying DQAS to both pools tests the generalizability of
the framework across this spectrum, from structured chemical pools to
fully hardware-native circuits, and provides a pathway to progressively
more compact ans\"{a}tze for near-term devices. As shown below, even
within the already gate-efficient QEB pool, circuits discovered by DQAS
remain shallower than those produced by ADAPT-VQE, demonstrating that
the gate-efficiency advantage stems from the global search strategy
itself rather than from properties of the UCCSD pool.

\begin{figure}[htbp]
  \centering
  \includegraphics[width=\linewidth]{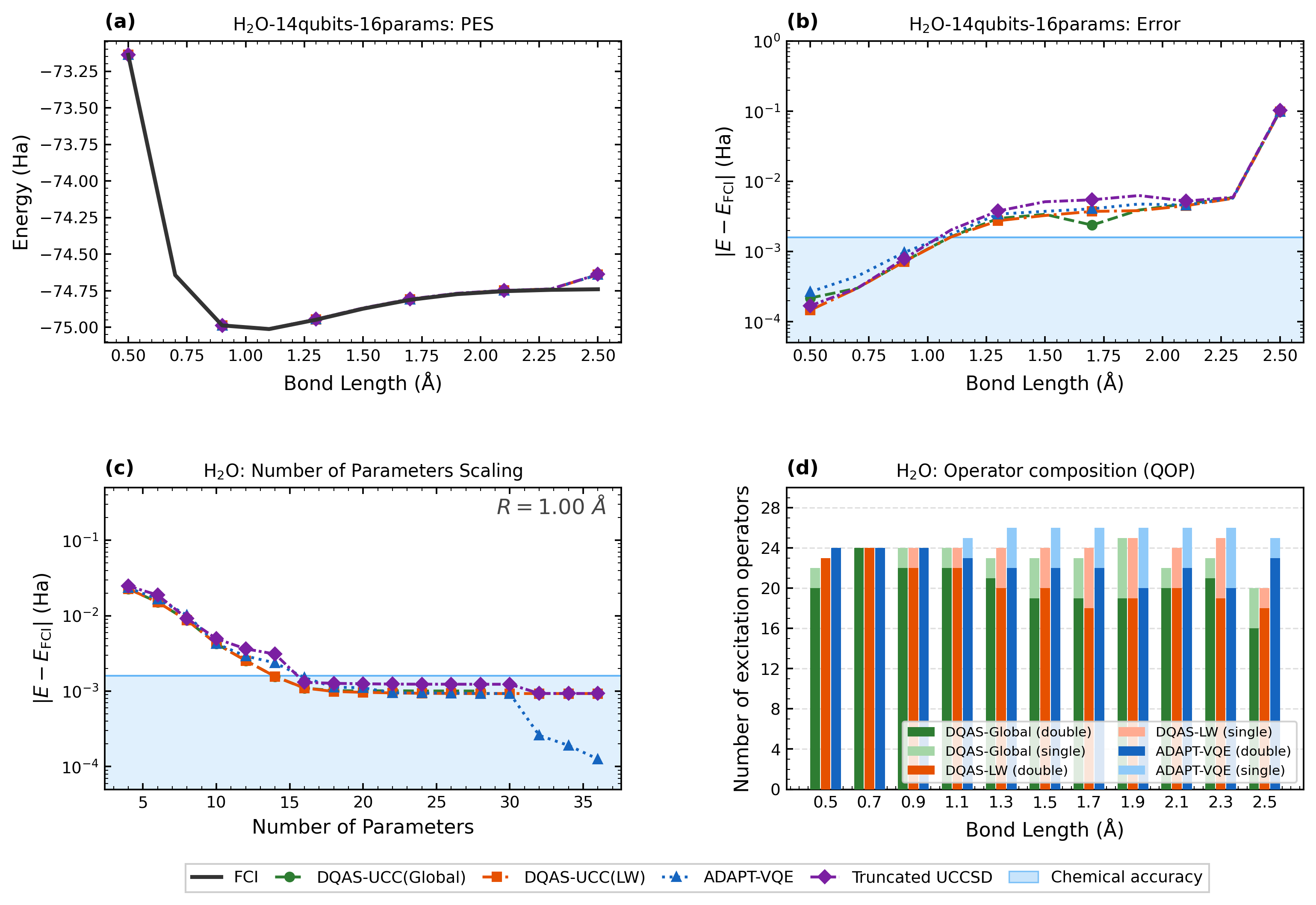}
  \caption{\textbf{DQAS with the qubit-excitation-based (QEB) operator pool for
  H$_2$O (14 qubits).}
  (a) Potential energy curve across $d = 0.5$--$2.5$ \AA.
  (b) Energy error versus bond length on a logarithmic scale.
  The shaded band denotes chemical accuracy ($<1.6$ mHa).
  (c) Energy error as a function of the number of variational parameters
  at $d = 1.00$ \AA; each QEB operator group contributes approximately
  two parameters.
  (d) Number of single and double excitation operators selected by each
  method at 16 operator groups across the dissociation profile; the QEB
  CNOT cost is 2 per single excitation and 14 per double excitation.}
  \label{fig:fig6}
\end{figure}

Figure~\ref{fig:fig6} presents the results of applying DQAS-Global
and DQAS-Layerwise with the QEB operator pool for H$_2$O, compared against
ADAPT-VQE and truncated UCCSD using the same qubit operator pool.
Panel (a) shows that both DQAS variants closely track the FCI
potential energy curve across the full dissociation profile
($d = 0.5$--$2.5$ \AA). Panel (b) confirms that both DQAS
variants satisfy the chemical accuracy threshold (errors
$<1.6$ mHa) at most bond lengths near equilibrium, with
DQAS-Layerwise achieving errors as low as 0.14 mHa at the
stretched geometry $d = 1.3$ \AA.

Panel (c) of Figure~\ref{fig:fig6} shows the energy error as a
function of the number of variational parameters at $d = 1.00$ \AA.
DQAS-Global (QEB) first reaches chemical accuracy at 14 operator groups
(28 parameters, error = 1.57 mHa), whereas QEB-ADAPT-VQE does not
reach chemical accuracy until 16 operator groups (32 parameters, error
= 1.52 mHa). At 16 groups, DQAS-Layerwise achieves 1.10 mHa versus
1.52 mHa for ADAPT-VQE, a 28\% improvement. This pattern mirrors the
UCCSD results and confirms that the DQAS advantage in parameter
efficiency is not restricted to a specific operator pool but is a
property of the global optimization strategy itself.

Panel (d) reveals the structural origin of the CNOT efficiency
advantage observed for the QEB pool. Because QEB single excitations
require only 2 CNOT gates and double excitations require 14, the
composition of the selected circuit directly determines the gate
count. Unlike UCCSD operators under the Jordan--Wigner
transformation, whose CNOT cost grows with qubit span, QEB operators
incur a fixed gate cost regardless of the spatial separation of the
involved orbitals, so the CNOT advantage in the QEB pool arises
entirely from operator-type composition rather than qubit-span
differences. Truncated UCCSD, which selects operators ranked by MP2
amplitude magnitude, consistently picks all double excitations at 16
operator groups (0 singles at most geometries), yielding 336 CNOT
gates per circuit. DQAS-Global and DQAS-Layerwise, by contrast, favor
a mix of single and double excitations: near equilibrium ($d =
0.9$--$1.7$ \AA), both DQAS variants select 2--6 single excitations
out of 22--25 total operators, reducing the CNOT count to 264--312
compared to 316--336 for ADAPT-VQE, a gate reduction of 4--17\%
within the QEB pool. This composition advantage is particularly visible
at stretched geometries ($d \geq 1.9$ \AA), where DQAS-Global selects
fewer total operators than ADAPT-VQE (22--25 vs 25--26 at these
geometries), amplifying the CNOT advantage.
This preference for single excitations is consistent with the
architecture--parameter coupling mechanism discussed in
Section~\ref{sec:cnot}: at the outset of training
($\boldsymbol{\theta} \approx \mathbf{0}$), single-excitation operators
already produce non-trivial energy gradients and immediately influence
the loss, causing the architecture parameters $\boldsymbol{\alpha}$ to
concentrate on them before double-excitation operators can compete on
equal footing.
Together, panels (c) and
(d) confirm that the DQAS gate-efficiency mechanism---global search
favoring cheaper operators over greedy gradient-guided selection---is
pool-independent and persists even within an already hardware-efficient
operator pool such as QEB.

As a further step toward fully hardware-native circuits, we also
applied DQAS to hardware-efficient ans\"{a}tze (HEA), where the gate
pool consists of generic single-qubit rotations ($R_x$, $R_y$, $R_z$)
and CNOT gates on adjacent qubit pairs without any chemical structure
imposed. HEA circuits are inherently more hardware-friendly than
UCC-based circuits: their gate sets align directly with native hardware
operations, and the unconstrained search space allows, in principle,
discovery of very compact entangling circuits tailored to a specific
device topology. Benchmarks on H$_2$ and LiH under both Jordan--Wigner
and parity encodings are reported in Figure~S1 of the
Supporting Information. These initial results indicate that applying
DQAS to HEA pools is substantially more challenging than applying it to
structured chemical pools: the absence of chemical prior information
forces the search to discover appropriate entanglement patterns entirely
from scratch, and the combinatorially larger search space increases
susceptibility to local optima. On the small systems studied
($\leq 6$ qubits), DQAS-HEA does not yet surpass fixed-topology
references such as RyLinear, reflecting the additional difficulty of
the unstructured setting rather than a fundamental limitation of the
DQAS framework itself.\cite{Belaloui2025}

Extending differentiable QAS to HEA pools on larger, more correlated
molecular systems remains a promising open direction.
Future approaches may benefit from incorporating hardware-specific noise
models directly into the DQAS objective, adopting hybrid pool designs
that combine hardware-native gate primitives with
physically motivated entanglement patterns, or using
physics-informed initialization strategies to reduce the effective search
space size.
We anticipate that combining the global optimization strategy of DQAS
with hardware-efficient circuit primitives will become increasingly
valuable as quantum devices scale and the gap between hardware-native
circuits and chemically structured ans\"{a}tze narrows.

% ===================================================================
\section{Conclusion}
% ===================================================================

Differentiable quantum architecture search offers an effective approach
to UCC circuit design by relaxing discrete operator selection into a
continuous differentiable optimization.
DQAS-Global and DQAS-Layerwise both achieve higher accuracy and fewer
CNOT gates than ADAPT-VQE in the compact-circuit regime.

Across the molecular systems studied, DQAS achieves up to 2.7-fold
accuracy improvement over ADAPT-VQE at equivalent operator counts,
while simultaneously requiring 13--17\% fewer CNOT gates for the UCCSD
pool.
The advantage is most pronounced at compact circuit depths; at larger
depths, gradient attenuation is a plausible contributing factor to
ADAPT-VQE's faster convergence, suggesting the two approaches are
complementary rather than universally ranked.
Benchmarks on the QEB operator pool confirm that both accuracy and
gate-efficiency advantages generalize beyond UCCSD.
These results demonstrate that differentiable architecture search
provides an effective and generalizable framework for designing
accurate and compact VQE circuits for molecular ground-state energy
calculations on near-term quantum devices.

% ===================================================================
%% Acknowledgments
% ===================================================================
\begin{acknowledgement}
The authors thank the High Performance Computing Center at The Chinese
University of Hong Kong, Shenzhen for computational resources.
The authors also thank Zhigang Shuai, Shixing Zhang, Yaxin Wang, and
Jinjun Zeng for helpful discussions.
This work was supported by the Shenzhen Science and Technology Program
(No.\ KQTD20240729102028011, JCYJ20250604140259001), the National
Natural Science Foundation of China (22573088), the Guangdong Basic
and Applied Basic Research Foundation (2026A1515010545), the Guangdong
Provincial Quantum Science Strategic Initiative
(Grant No.\ GDZX2503001), and the Guangdong Basic Research Center of
Excellence for Aggregate Science.
\end{acknowledgement}

% ===================================================================
%% References
% ===================================================================
\bibliography{references}

\end{document}